\documentclass[9pt,twocolumn,twoside]{opticajnl}
\journal{opticajournal} % use for journal or Optica Open submissions

\setboolean{shortarticle}{true}
\usepackage{lineno}
\title{Reducing the Carrier-Envelope-Phase-dependence of High-Harmonic-Generation by Vectorial-Time-Polarization-Gating}

\author[1,2]{Eran Ben-Arosh}
\author[1,2]{Eldar Ragonis}
\author[1,2]{Lev Merensky}
\author[1,2,*]{Avner Fleischer}

\affil[1]{Raymond and Beverly Sackler Faculty of Exact Sciences, School of Chemistry, Tel Aviv University, Tel Aviv 6997801, Israel}
\affil[2]{Tel-Aviv University Center for Light-Matter-Interaction, Tel Aviv 6997801, Israel}

\affil[*]{avnerfleisch@tauex.tau.ac.il}

\begin{abstract}
A well-known shortcoming of High Harmonic Generation (HHG) is the strong dependence of the broadband HHG spectra (HGS) on the carrier envelope phase (CEP) of the driver. Here we numerically show that compared to the current well-established scalar (linearly polarized) schemes for generating broadband HGS, namely a short driver [Amplitude gating (AG)], Polarization-Gating (PG) or Time-Gating (TG), the vectorial driver of the Vectorial-Time-Polarization-Gating (VTPG) scheme renders the cutoff HGS much less sensitive to the CEP of the driver. The polarization state (helicity) of the emitted radiation is likewise CEP-resilient. Unlike scalar schemes, where the number of recollisions heavily depends on the CEP, in VTPG the CEP keeps this number almost unchanged, and only controls the partitioning of the recollisions between two orthogonal directions. This reduces the CEP-dependence of the HGS and decreases the spectral modulations. The CEP-resilience of the VTPG scheme holds promise for a variety of applications in attosecond science benefiting from quasi-continuous, helical HHG sources liberated from the necessity to stabilize the CEP of the laser.
\end{abstract}

\setboolean{displaycopyright}{false} % Do not include copyright or licensing information in submission.

\begin{document}

\maketitle

\begin{figure}[!b]
\centering
\includegraphics[width=\linewidth]{plots/Figure_33.png}
\caption{The Vectorial-Time-Polarization-Gating (VTPG) scheme with cross-linear two-color driver: gate separation and duration dependence on the instantaneous ellipticity of the driver field. $\mathbf{(a)}$ Lissajous curve of the VTPG electric field with $\delta=0.05$. $\mathbf{(b,c,d)}$ show the absolute value of the driver field (left vertical axis) for $\delta=0.025,0.05,0.1$ respectively as well as their instantaneous ellipticity (black lines; right vertical axis) with the dashed horizontal line indicating a critical ellipticity value of $\varepsilon^{c} = 0.2$. Note that $\mathbf{(a)}$ and $\mathbf{(c)}$ describe the same field: it obtains very low instantaneous ellipticities at times 0, 5T, 10T, 15T, 20T. At those instants recollisions can take place and emission bursts are obtained. $\mathbf{(e)}$ shows the instantaneous ellipticity map as a function of the detuning factor and time. Increasing the detuning factor shortens the gate duration and reduces the time separation between consecutive gates.}
\label{fig:ellipticity_MAP_analysis}
\end{figure}

\noindent\textbf{Introduction.}
In recent decades, High Harmonic Generation (HHG) has become the most widely used technique to generate coherent extreme-ultraviolet (XUV) radiation from a table-top source \cite{corkum2007attosecond,krausz2009attosecond,biegert2021attosecond,alexander2025attosecond}. Although the typically sparse HHG spectrum (HGS) is valuable for many applications, other applications such as Attosecond Transient Absorption Spectroscopy (ATAS) require sources delivering broadband XUV radiation \cite{geneaux2019transient,di2024attosecond}. To that end, the nature of the HHG process requires reducing the number of recollision events, as this results in the generation of a broader spectrum \cite{christov1997high}. A straightforward method to achieve this is to simply shorten the duration of the driver ("Amplitude Gating" [AG]). This is usually achieved by spectral broadening techniques, such as focusing the laser to a gas-filled hollow-core fiber (HCF) followed by chirp compensation \cite{nisoli1996generation}. Other gating methods bypass the need for post-compression of the driver pulse by confining the HHG generation process to short temporal instances ("gates"). Common examples are the "Polarization-Gating" (PG) \cite{corkum1994subfemtosecond, shan2005generation} and "Time Gating" (TG) \cite{fleischer2006attosecond,merdji2007isolated} techniques. In all methods, as the effective gate duration approaches the regime of a single optical cycle \cite{baltuvska2003attosecond,kienberger2004atomic}, Carrier-Envelope-Phase (CEP, hereby defined as $\phi$) becomes an important parameter as it greatly affects the number of recollisions obtained, ranging from a single recollision (when the field forms a "cos" pulse, where the carrier and envelope are in phase) to two recollisions ("sin" pulse, when the CEP is 0.5$\pi$) or any intermediate number. Hence, shot to shot changes in CEP will lead to large spectral variations \cite{chini2014generation, nisoli2003effects,frolov2012analytic}. Common approaches to mitigate CEP sensitivity include the use of CEP-stable driver lasers, which are constrained by the limited response of the CEP feedback mechanism, or by characterizing the CEP in each shot by CEP-tagging \cite{baltuska2003phase,cundiff2002phase,xue2021custom,gollner2025carrier}, and binning the obtained experimental data, possibly discarding that which did not originate from driver pulses within the desired CEP range. Although some applications could utilize self-selection of an isolated attosecond pulse obtained for a particular range of CEP values \cite{gilbertson2010isolated}, albeit at the expense of large shot-to-shot fluctuations of the HGS flux, minimizing such fluctuations and obtaining true CEP-invariant broadband HHG schemes are important for many applications in attosecond science. Moreover, the schemes mentioned above not only suffer from CEP-sensitivity but are also only capable of generating close-to-linear polarization of the HHG radiation. Some applications, such as the detection of chiral matter, would however benefit from XUV radiation whose polarization state is highly-elliptical (possibly circular).

\begin{figure*}[!t]
\centering
\includegraphics[width=\textwidth]{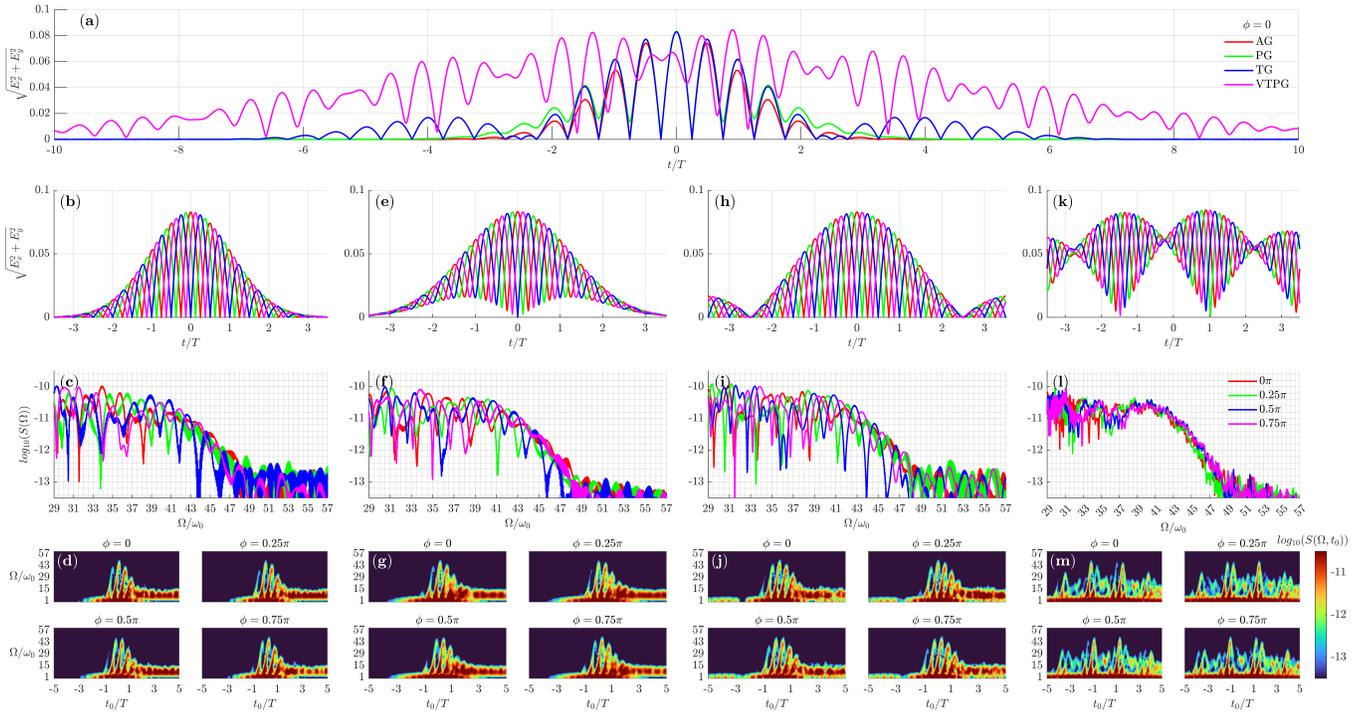}
\caption{Comparison of HHG spectra and their Gabor transforms at 4 CEP values, $\phi = 0, 0.25\pi, 0.5\pi, 0.75\pi$ (red, green, blue and magenta, respectively) produced via standard schemes and the VTPG 2-gate, 3-recollision case. $\mathbf{a}$: absolute field values of AG, TG, PG and VTPG for $\phi$ = 0. $\mathbf{(b,c,d)}$: absolute field, spectrum and Gabor transform of AG, a monochromatic linearly polarized short pulse. $\mathbf{(e,f,g)}$: same, but for PG. $\mathbf{(h,i,j)}$: same, but for TG. $\mathbf{(k,l,m)}$: same, but for the VTPG 2-gate, 3-recollision case. Note how the CEP-dependence and the visibility (modulation depth) are greatly reduced.}
\label{fig:CEP_spectrum_analysis}
\end{figure*}

Here we show that the Vectorial Time Polarization-Gating (VTPG) scheme \cite{ragonis2024controlling,benArosh2025Chiral}, when appropriately tuned, greatly reduces the CEP sensitivity of the HGS. This is achieved because unlike the scalar schemes mentioned above (AG, PG, TG), in VTPG the number of recollisions remains nearly invariant to the value of the CEP. The CEP only affects their internal distribution between two orthogonal directions. The VTPG scheme is based on a two-color source with slightly detuned frequencies, where in this work, the two-colors are linearly-polarized along orthogonal directions ("cross-linear" case). This gating technique restricts the recollisions to well-defined temporal windows or "gates", at which the driver is close to being linearly-polarized and hence intense recollisions are obtained. The adjacent gates are perpendicularly-oriented and are related in terms of the carrier field \cite{benArosh2025Chiral}. For instance, if the first gate supports a "sin" field, then the following gate would support a "cos" field. This property fixes the total number of recollisions within two adjacent gates, regardless of the value of the CEP. By analyzing the instantaneous ellipticity of the driver \cite{benArosh2025Chiral},
we find that the duration of the gate is approximately given by
$\tau_{\mathrm{gate}} \approx \frac{\varepsilon^{c} T}{\pi \delta}$ where $\varepsilon^{c}$ is the critical ellipticity above which recollision is suppressed (commonly taken as $\varepsilon=0.2$ \cite{antoine1996theory,shan2005generation}), $T = \frac{2 \pi}{\omega_{0}}$ ($\omega_{0}$ is the central frequency of the driving field) and $\delta$ is the frequency detuning parameter, given by $\delta = \frac{\omega_{1}-\omega_{2}}{2\omega_{0}}$, where $\omega_{1} = \omega_{0}(1 + \delta),~\omega_{2} = \omega_{0}(1 - \delta)$ are the frequencies of the two colors, respectively. Fig. \ref{fig:ellipticity_MAP_analysis} shows that as $\delta$ decreases, the gates become longer, the transition to and out of the linear polarization state occurs more gradually, and the temporal separation between consecutive gates increases. In contrast, increasing $\delta$ results in shorter gates, faster transitions, and reduced separation between gates. For $\delta$ = 0.1 (Fig. \ref{fig:ellipticity_MAP_analysis}b) the carrier during the gate at t = 10T does not reach zero, that is, this gate represents a "cos" pulse. However, the carrier at the next gate (at t = 12.5T) does reach zero, that is, this gate represents a "sin" pulse. The key physical mechanism underlying CEP insensitivity in the VTPG scheme is this relation between every two consecutive gates which results in the preservation of the total number of recollisions, provided the total number of gates is even with the minimal number of gates required being two. In a sense, the first emission burst is complemented with a second burst in order to maintain the total number of recollisions roughly fixed and CEP-independent.

\noindent\textbf{Methods.}
We numerically compare the "conventional" (i.e., scalar) approaches for generating quasi-continuous HGS to a particular case of the VTPG scheme in which the two-color carrier field is cross-linearly polarized, and the total wavefront has the minimal number of gates (two) and minimal number of recollisions per gate (minimum 1, maximum 2). We term this waveform the "2-gate, 3-recollision" field. We consider a standard output of a femtosecond Ti:Sa laser, with a spectrum centered at $800$ nm, detuned by $\delta = 0.1$. The maximal instantaneous amplitude is $E_{max} = \sqrt{2} \cdot0.058$, corresponding to a cutoff at the $39^{\mathrm{th}}$ harmonic. For each scheme, we solve the three-dimensional time-dependent Schrödinger equation (TDSE) within a single-active-electron (SAE) model on a Cartesian grid using the split-operator method \cite{feit1982solution}. We employ atomic units, $\hbar = m = -e = 1$, and treat the interaction with the laser within the length gauge:
\begin{equation}
i\,\frac{\partial}{\partial t}\Psi(\mathbf{r},t)
=
\left[
-\frac{1}{2}\nabla^{2}
+ V(\mathbf{r})
+ \mathbf{E}(t) \cdot \mathbf{r}
\right]\Psi(\mathbf{r},t),
\label{eq:TDSE}
\end{equation}
where the model potential
\(
V(\mathbf{r})
=
-\frac{1 + A\,e^{-B\mathbf{r}}}{\sqrt{C + \mathbf{r}^{2}}}
\)
(\(A=0.2719\), \(B=0.25\) and \(C=0.09192\)), reproduces the ionization potential of argon~\cite{fleischer2017polarization}. Using Ehrenfest’s theorem, we calculate the dipole acceleration, \(\mathbf{a}(t)=\Big\langle \Psi(\mathbf{r},t)\,\Big|\, -\nabla V(\mathbf{r})-\mathbf{E}(t)\,\Big|\,\Psi(\mathbf{r},t)\Big\rangle\). The HGS is obtained by Fourier transforming each Cartesian component and summing the corresponding spectral intensities $S(\Omega)\propto |\overline{a}_x(\Omega)|^2+|\overline{a}_y(\Omega)|^2$, where $\overline{a}_j(\Omega)=\int_{-\infty}^{\infty} a_j(t)\,e^{-i\Omega t}\,dt$ and $j\in\{x,y\}$ denotes the Cartesian components of the vectors in the polarization plane. We define the general driving laser field as
\begin{equation}
\begin{aligned}
\mathbf{E}(t)
&=
E_{0} f_{1}(t - t_{1})
\begin{bmatrix}
\eta_{1x}\cos\!\left(\omega_{1}(t-t_{1}) + \phi\right) \\
\eta_{1y}\sin\!\left(\omega_{1}(t-t_{1}) + \phi \right)
\end{bmatrix}
\\
&\quad+
E_{0} f_{2}(t - t_{2})
\begin{bmatrix}
\eta_{2x}\cos\!\left(\omega_{2}(t-t_{2}) + \phi \right) \\
\eta_{2y}\sin\!\left(\omega_{2}(t-t_{2}) + \phi \right)
\end{bmatrix},
\end{aligned}
\label{eq:Efield_twocolor_vector_centers}
\end{equation}
where the square parentheses notation indicates the components in the polarization plane, $f_k(t-t_k)=\exp[-2\ln 2\,((t-t_k)/t_p)^2]$, $k\in\{1,2\}$ are Gaussian envelopes, and $\phi$ is the CEP. Here $t_{1}$ and $t_{2}$ denote the temporal centers of the two color components, and $t_p$ is the full width at half maximum (FWHM) duration of the intensity envelope. The parameters $\eta_{j\alpha}\in\{-1,0,1\}$ determine whether the component $\alpha\in\{x,y\}$ of the field $j$ is present and its sign. The benchmark schemes are defined as follows:
\begin{equation*}
\begin{aligned}
\text{AG:}\quad
& \eta_{1x}=1,\ \eta_{1y}=\eta_{2x}=\eta_{2y}=0, \\
& \omega_{1}=\omega_{2}=\omega_{0}, \\
& E_{0}=\sqrt{2} \cdot 0.058\ ,\ \tau_{p}=1.75T,\ t_{1}=t_{2}=0,
\\
\text{PG:}\quad
& \eta_{1x}=\ \eta_{1y}=\ \eta_{2x}=1,\ \eta_{2y}=-1, \\
& \omega_{1}=\omega_{2}=\omega_{0}, \\
& E_{0}=0.044\ ,\ \tau_{p}=2T,\ t_{1}=-t_{2}=-0.75T,
\\
\text{TG:}\quad
& \eta_{1x}=\eta_{2x}=1,\ \eta_{1y}=\eta_{2y}=0, \\
& \omega_{1}=\omega_{0}(1+\delta),\ \omega_{2}=\omega_{0}(1-\delta),\ \delta=0.1, \\
& E_{0}=0.041\ ,\ \tau_{p}=4T,\ t_{1}=t_{2}=0,
\\
\text{VTPG:}\quad
& \eta_{1x}=0,\ \eta_{1y}=\eta_{2x}=1,\ \eta_{2y}=0, \\
& \omega_{1}=\omega_{0}(1+\delta),\ \omega_{2}=\omega_{0}(1-\delta),\ \delta=0.1, \\
& E_{0}=0.058\ ,\ \tau_{p}=8T,\ t_{1}=-t_{2}=-0.55T.
\end{aligned}
\label{eq:fields_4schemes_template}
\end{equation*}

To resolve the temporal evolution of the spectral components, a time-frequency analysis is performed by means of a Gabor transform, which is a Fourier transform of the dipole acceleration components evaluated within a sliding Gaussian time window, 
$G_j(t_0,\Omega)=\int_{-\infty}^{\infty} a_j(t)\,
\exp\!\left[-2\ln2\left(\frac{t-t_0}{\tau_g}\right)^2\right]\,
e^{-i\Omega t}\,dt$, where $t_0$ and $\tau_{g} = 0.1T$ are the window's center and width, respectively. Note that for the central frequency of $800$ nm (optical period of $\approx2.669$ fs), spectral detuning $\delta = 0.1$ and critical ellipticity of $0.2$, the duration of each gate is $1.699$ fs and adjacent gates are separated by $\frac{T}{4\delta} = 6.671$ fs. The two initial pulses considered here each have a duration of \(21.34\) fs. Hence, the VTPG technique reduces the number of recollisions by a factor of 13 with respect to its linear pulse constituents. 

\begin{figure}[!t]
\centering
\includegraphics[width=\columnwidth]{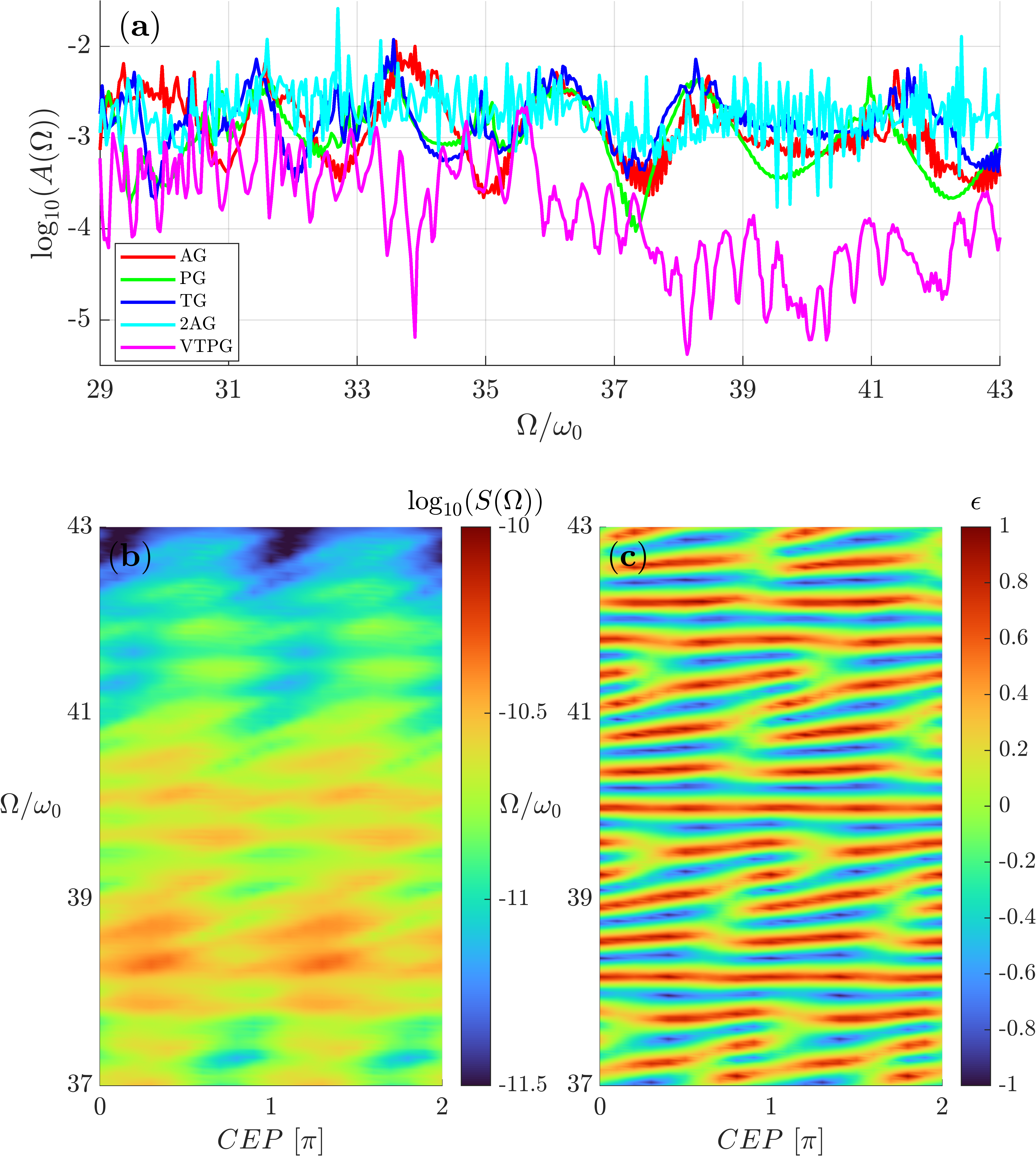}
\caption{$\mathbf{(a).}$ CEP-sensitivity metric $A(\Omega)$ of the 5 methods. HGS $\mathbf{(b)}$ and the ellipticity $\varepsilon$ $\mathbf{(c)}$ vs. CEP for the
VTPG 2-gate, 3-recollision case.}
\label{fig:CEP_MAPS_analysis}
\end{figure}

\noindent\textbf{Results and discussion.}
Fig. \ref{fig:CEP_spectrum_analysis} and Fig. \ref{fig:CEP_MAPS_analysis}a compare the HGS obtained for the four methods. In all three methods for which the driver is scalar (AG, PG, TG), the HGS is strongly dependent on the value of the CEP since this value greatly affects the number of recollisions, which here (in the cutoff region) varies between 2 and 3. The VTPG scheme is designed to support 2 gates (centered around t = -1.25T and t = 1.25T) as this is the minimal number of gates which yields a vectorial driver. Depending on the value of the CEP, the total number of recollisions is again between 2 and 3, however, unlike in the scalar schemes not all are pointing in the same direction. In the VTPG scheme, the time-frequency analysis reveals a monotonic dependence of the recollision waveform on the CEP, as a result of the gate relation described before. As the CEP is gradually increased, a recollision event from the leading edge of the first gate is removed, while a new recollision at the trailing edge of the next gate appears. Different CEP values only cause a redistribution of the recollisions between the orthogonally polarized gates but the total number of recollisions is roughly kept constant, in contrast to the linearly polarized scalar schemes where the CEP greatly affects the number of recollisions. The HGS, however, depends not only on the total number of recollisions but more importantly on the exact emission sequence, which in general varies with the CEP. This is because the emitted field is a result of the coherent interference of all emission events. The crucial advantage of VTPG is that the two gates are polarized along perpendicular directions; as shown in Fig.~\ref{fig:ellipticity_MAP_analysis}a, the gate at \(t=10T\) points at \(45^\circ\) to the x-axis, whereas the following gate at \(t=15T\) points at \(-45^\circ\). In fact, the perpendicular polarizations of successive gates ensure that spectrum-wise the emission burst from the first gate does \textbf{not} interfere with that of the second gate as perpendicular fields do not interfere. However, field intensities (rather than amplitudes) do add up. If 2 recollisions are induced in the first gate, (as occurs for $\phi = 0.25\pi$) quasi-continuous emission with noticeable spectral modulations (with maxima at odd and minima at even harmonic orders) is expected. These modulations are mitigated by the continuous-emission generated by the single recollision which occurs in the subsequent gate. Were this single recollision the only emission event, it would have resulted in HGS with no modulations. Hence, the emission from the second gate contributes a pedestal (baseline) which reduces the modulation depth (HGS visibility) of the total HGS signal relative to the HGS generated by the first gate alone. To quantify the CEP sensitivity of the HGS, we define the CEP-dependent variable \(X(\Omega,\phi_{n})=\log_{10} S(\Omega,\phi_{n})\) and its CEP-average
\(
\bar{X}(\Omega)=\frac{1}{N}\sum_{n=1}^{N} X(\Omega,\phi_{n}).
\)
We then introduce the normalized variance metric \(A(\Omega)\),
\begin{equation}
A(\Omega)=
\frac{1}{N}
\sum_{n=1}^{N}
\frac{\left[X(\Omega,\phi_{n})-\bar{X}(\Omega)\right]^2}
{|\bar{X}(\Omega)|^2},
\label{eq:CEP_metric}
\end{equation}
and calculate \(A(\Omega)\) for all methods using 20 equidistant values of \(\phi\) between \(0\) and \(2\pi\), i.e., \(N = 20\). Smaller values of \(A(\Omega)\) correspond to improved CEP insensitivity. Fig. \ref{fig:CEP_MAPS_analysis}a compares this quantity for the different schemes discussed above where the superiority of the VTPG scheme over the scalar schemes is clearly evident. A fifth scalar scheme, denoted '2AG', is also considered, in which HHG is driven by a carrier structured into two short linearly polarized pulses separated by \(4.75T\). One may reasonably argue that this configuration also preserves a constant total number of recollisions, since a cosine-waveform generated by the first pulse is necessarily followed by a sine-waveform in the second pulse. However, its performance is comparable to the other scalar schemes, showing that rotating the second emission burst orthogonally to the first (as in VTPG) is essential for CEP stability.

Finally, we analyze the behavior of the harmonic polarization states. Fig.~\ref{fig:CEP_MAPS_analysis}c shows that the ellipticity of the emitted harmonics is stable with respect to the CEP, although to a lesser extent than the spectrum. Since polarization is a vector quantity, emissions along perpendicular directions do affect one another and jointly determine the total polarization state at each XUV frequency. The polarization states were calculated according to:
$\varepsilon(\Omega)=\sqrt{\frac{1-\eta(\Omega)}{1+\eta(\Omega)}}$, with $\eta(\Omega)\equiv \frac{1}{S(\Omega)}\sqrt{S^{2}(\Omega)-4\,|\overline{a}_{x}(\Omega)|^{2}\,|\overline{a}_{y}(\Omega)|^{2}\sin^{2}\!\bigl[\phi_{y}(\Omega)-\phi_{x}(\Omega)\bigr]}$
where \(\varepsilon\) is the XUV ellipticity, and \(\phi_x\) and \(\phi_y\) are the phases of \(a_x(t)\) and \(a_y(t)\), respectively \cite{benArosh2025Chiral}.

\noindent\textbf{Conclusion.}
%\section*{Conclusion}
We presented a high-harmonic generation scheme based on a vectorial driving field that naturally suppresses sensitivity to CEP fluctuations. The use of slightly detuned, two-color, cross-linear field components structures the electron recollision dynamics into temporally separated orthogonal gates. Variations in the CEP modify the timing and emission directions of individual recollision events but leave their overall count unchanged, resulting in a stable harmonic response. A time–frequency analysis shows that recollisions at adjacent gates are related, while the orthogonality of successive gates prevents cross-interference. This mechanism preserves a broadband spectral region near and beyond the cutoff, even for multi-cycle driving pulses. In addition, the highly elliptical polarization state of the emitted harmonics also remains stable under CEP variation. These characteristics establish the VTPG scheme as a robust platform for broadband attosecond and chiral spectroscopy, particularly in experimental configurations where short-pulse operation and active CEP stabilization are impractical.

\begin{backmatter}
%\bmsection{Funding}

%\bmsection{Acknowledgment} .

\bmsection{Disclosures} The authors declare no conflicts of interest.

%\smallskip

\bmsection{Data availability} Data underlying the results presented in this paper may be obtained from the authors upon reasonable request.

\end{backmatter}

% Bibliography
\bibliography{sample}

@article{corkum2007attosecond,
  title={Attosecond science},
  author={Corkum, Paul B and Krausz, Ferenc},
  journal={Nature physics},
  volume={3},
  number={6},
  pages={381--387},
  year={2007},
  publisher={Nature Publishing Group UK London}
}

@article{biegert2021attosecond,
  title={Attosecond technology (ies) and science},
  author={Biegert, Jens and Calegari, Francesca and Dudovich, Nirit and Qu{\'e}r{\'e}, Fabien and Vrakking, Marc},
  journal={Journal of Physics B: Atomic, Molecular and Optical Physics},
  volume={54},
  number={7},
  pages={070201},
  year={2021},
  publisher={IOP Publishing}
}

@article{alexander2025attosecond,
  title={Attosecond physics and technology},
  author={Alexander, O and Ayuso, D and Matthews, M and Rego, L and Tisch, JWG and Weaver, B and Marangos, JP},
  journal={Applied Physics Letters},
  volume={126},
  number={17},
  year={2025},
  publisher={AIP Publishing}
}

@article{geneaux2019transient,
  title={Transient absorption spectroscopy using high harmonic generation: a review of ultrafast X-ray dynamics in molecules and solids},
  author={Geneaux, Romain and Marroux, Hugo JB and Guggenmos, Alexander and Neumark, Daniel M and Leone, Stephen R},
  journal={Philosophical Transactions of the Royal Society A},
  volume={377},
  number={2145},
  pages={20170463},
  year={2019},
  publisher={The Royal Society Publishing}
}

@article{di2024attosecond,
  title={Attosecond absorption and reflection spectroscopy of solids},
  author={Di Palo, Nicola and Inzani, Giacomo and Dolso, GL and Talarico, Matteo and Bonetti, Simone and Lucchini, Matteo},
  journal={APL Photonics},
  volume={9},
  number={2},
  year={2024},
  publisher={AIP Publishing}
}

@article{christov1997high,
  title={High-harmonic generation of attosecond pulses in the “single-cycle” regime},
  author={Christov, Ivan P and Murnane, Margaret M and Kapteyn, Henry C},
  journal={Physical Review Letters},
  volume={78},
  number={7},
  pages={1251},
  year={1997},
  publisher={APS}
}

@article{kienberger2004atomic,
  title={Atomic transient recorder},
  author={Kienberger, Reinhard and Goulielmakis, Eleftherios and Uiberacker, Matthias and Baltuska, Andrius and Yakovlev, Vladislav and Bammer, Ferdinand and Scrinzi, Armin and Westerwalbesloh, Th and Kleineberg, Ulf and Heinzmann, Ulrich and others},
  journal={Nature},
  volume={427},
  number={6977},
  pages={817--821},
  year={2004},
  publisher={Nature Publishing Group UK London}
}

@article{baltuvska2003attosecond,
  title={Attosecond control of electronic processes by intense light fields},
  author={Baltu{\v{s}}ka, Andrius and Udem, Th and Uiberacker, Matthias and Hentschel, Michael and Goulielmakis, Eleftherios and Gohle, Ch and Holzwarth, Ronald and Yakovlev, Vladislav S and Scrinzi, Armin and H{\"a}nsch, Th W and others},
  journal={Nature},
  volume={421},
  number={6923},
  pages={611--615},
  year={2003},
  publisher={Nature Publishing Group UK London}
}

@article{nisoli2003effects,
  title={Effects of Carrier-Envelope Phase Differences of Few-Optical-Cycle Light Pulses<? format?> in Single-Shot High-Order-Harmonic Spectra},
  author={Nisoli, Mauro and Sansone, Giuseppe and Stagira, Salvatore and De Silvestri, Sandro and Vozzi, Caterina and Pascolini, M and Poletto, L and Villoresi, Paolo and Tondello, Giuseppe},
  journal={Physical review letters},
  volume={91},
  number={21},
  pages={213905},
  year={2003},
  publisher={APS}
}

@article{cundiff2002phase,
  title={Phase stabilization of ultrashort optical pulses},
  author={Cundiff, Steven T},
  journal={Journal of Physics D: Applied Physics},
  volume={35},
  number={8},
  pages={R43},
  year={2002},
  publisher={IOP Publishing}
}

@article{xue2021custom,
  title={A custom-tailored multi-TW optical electric field for gigawatt soft-X-ray isolated attosecond pulses},
  author={Xue, Bing and Tamaru, Yuuki and Fu, Yuxi and Yuan, Hua and Lan, Pengfei and M{\"u}cke, Oliver D and Suda, Akira and Midorikawa, Katsumi and Takahashi, Eiji J},
  journal={Ultrafast Science},
  year={2021},
  publisher={AAAS}
}

@article{gollner2025carrier,
  title={Carrier-envelope-phase characterization of ultrafast mid-infrared laser pulses through harmonic generation and interference in argon},
  author={Gollner, Claudia and Shumakova, Valentina and Barker, Jacob and Pug{\v{z}}lys, Audrius and Baltu{\v{s}}ka, Andrius and Polynkin, Pavel},
  journal={Communications Physics},
  volume={8},
  number={1},
  pages={33},
  year={2025},
  publisher={Nature Publishing Group UK London}
}

@article{ragonis2024controlling,
  title={Controlling the bandwidth of high harmonic emission peaks with the spectral polarization of the driver},
  author={Ragonis, Eldar and Ben-Arosh, Eran and Merensky, Lev and Fleischer, Avner},
  journal={Optics Letters},
  volume={49},
  number={10},
  pages={2741--2744},
  year={2024},
  publisher={Optica Publishing Group}
}

@article{benArosh2025Chiral,
    author = {Ben-Arosh, Eran and Ragonis, Eldar and Merensky, Lev and Fleischer, Avner},
    title = {Chiral broadband High Harmonic Generation Source by Vectorial Time-Polarization-Gating},
    journal= {arXiv:2503.03970},
    eprinttype = {arxiv},
    year = {2025},
    url = {https://arxiv.org/abs/2503.03970},
    urldate = {2025-03-05}
}

@article{fleischer2017polarization,
  title={Polarization-fan high-order harmonics},
  author={Fleischer, Avner and Bordo, Eliyahu and Kfir, Ofer and Sidorenko, Pavel and Cohen, Oren},
  journal={Journal of Physics B: Atomic, Molecular and Optical Physics},
  volume={50},
  number={3},
  pages={034001},
  year={2017},
  publisher={IOP Publishing}
}

@article{nisoli1996generation,
  title={Generation of high energy 10 fs pulses by a new pulse compression technique},
  author={Nisoli, Mauro and De Silvestri, Sandro and Svelto, Orazio},
  journal={Applied Physics Letters},
  volume={68},
  number={20},
  pages={2793--2795},
  year={1996},
  publisher={American Institute of Physics}
}

@article{chini2014generation,
  title={The generation, characterization and applications of broadband isolated attosecond pulses},
  author={Chini, Michael and Zhao, Kun and Chang, Zenghu},
  journal={Nature Photonics},
  volume={8},
  number={3},
  pages={178--186},
  year={2014},
  publisher={Nature Publishing Group UK London}
}

@article{frolov2012analytic,
  title={Analytic theory of high-order-harmonic generation by an intense few-cycle laser pulse},
  author={Frolov, MV and Manakov, NL and Popov, AM and Tikhonova, OV and Volkova, EA and Silaev, AA and Vvedenskii, NV and Starace, Anthony F},
  journal = {Physical Review A: Atomic, Molecular, and Optical Physics},
  volume={85},
  number={3},
  pages={033416},
  year={2012},
  publisher={APS}
}

@article{gilbertson2010isolated,
  title={Isolated Attosecond Pulse Generation without the Need to Stabilize the Carrier-Envelope Phase<? format?> of Driving Lasers},
  author={Gilbertson, Steve and Khan, Sabih D and Wu, Yi and Chini, Michael and Chang, Zenghu},
  journal={Physical review letters},
  volume={105},
  number={9},
  pages={093902},
  year={2010},
  publisher={APS}
}

@article{baltuska2003phase,
  title={Phase-controlled amplification of few-cycle laser pulses},
  author={Baltuska, Andrius and Uiberacker, Matthias and Goulielmakis, Eleftherios and Kienberger, Reinhard and Yakovlev, Vladislav S and Udem, Thomas and Hansch, Theodor W and Krausz, Ferenc},
  journal={IEEE Journal of Selected Topics in Quantum Electronics},
  volume={9},
  number={4},
  pages={972--989},
  year={2003},
  publisher={IEEE}
}

@article{shan2005generation,
  title={Generation of the attosecond extreme ultraviolet supercontinuum by a polarization gating},
  author={Shan, Bing and Ghimire, Shambhu and Chang, Zenghu},
  journal={Journal of modern optics},
  volume={52},
  number={2-3},
  pages={277--283},
  year={2005},
  publisher={Taylor \& Francis}
}

@article{corkum1994subfemtosecond,
  title={Subfemtosecond pulses},
  author={Corkum, PB and Burnett, NH and Ivanov, M Yu},
  journal={Optics letters},
  volume={19},
  number={22},
  pages={1870--1872},
  year={1994},
  publisher={Optical Society of America}
}

@article{merdji2007isolated,
  title={Isolated attosecond pulses using a detuned second-harmonic field},
  author={Merdji, Hamed and Auguste, Thierry and Boutu, Willem and Caumes, J-Pascal and Carr{\'e}, Bertrand and Pfeifer, Thomas and Jullien, Aur{\'e}lie and Neumark, Daniel M and Leone, Stephen R},
  journal={Optics letters},
  volume={32},
  number={21},
  pages={3134--3136},
  year={2007},
  publisher={Optical Society of America}
}

@article{fleischer2006attosecond,
  title={Attosecond laser pulse synthesis using bichromatic high-order harmonic generation},
  author={Fleischer, Avner and Moiseyev, Nimrod},
  journal={Physical Review A},
  volume={74},
  number={5},
  pages={053806},
  year={2006},
  publisher={APS}
}

@article{krausz2009attosecond,
  title={Attosecond physics},
  author={Krausz, Ferenc and Ivanov, Misha},
  journal={Reviews of modern physics},
  volume={81},
  number={1},
  pages={163--234},
  year={2009},
  publisher={APS}
}

@article{antoine1996theory,
  title={Theory of high-order harmonic generation by an elliptically polarized laser field},
  author={Antoine, Philippe and L’Huillier, Anne and Lewenstein, Maciej and Sali{\`e}res, Pascal and Carr{\'e}, Bertrand},
  journal={Physical Review A},
  volume={53},
  number={3},
  pages={1725},
  year={1996},
  publisher={APS}
}

@article{feit1982solution,
  title={Solution of the Schr{\"o}dinger equation by a spectral method},
  author={Feit, MD and Fleck Jr, JA and Steiger, A},
  journal={Journal of Computational Physics},
  volume={47},
  number={3},
  pages={412--433},
  year={1982},
  publisher={Elsevier}
}

% Please include bios and photos of all authors for aop articles
\ifthenelse{\equal{\journalref}{aop}}{%
\section*{Author Biographies}
\begingroup
\setlength\intextsep{0pt}
\begin{minipage}[t][6.3cm][t]{1.0\textwidth} % Adjust height [6.3cm] as required for separation of bio photos.
  \begin{wrapfigure}{L}{0.25\textwidth}
    \includegraphics[width=0.25\textwidth]{john_smith.eps}
  \end{wrapfigure}
  \noindent
  {\bfseries John Smith} received his BSc (Mathematics) in 2000 from The University of Maryland. His research interests include lasers and optics.
\end{minipage}
\begin{minipage}{1.0\textwidth}
  \begin{wrapfigure}{L}{0.25\textwidth}
    \includegraphics[width=0.25\textwidth]{alice_smith.eps}
  \end{wrapfigure}
  \noindent
  {\bfseries Alice Smith} also received her BSc (Mathematics) in 2000 from The University of Maryland. Her research interests also include lasers and optics.
\end{minipage}
\endgroup
}{}

\end{document}